\documentclass[preprint,sort&compress,times,fleqn]{elsarticle}
\usepackage{slashed}
\usepackage{amssymb}
\usepackage{amsmath}
\usepackage{amsfonts}
\usepackage{epsfig}
\usepackage{graphicx}
\usepackage{tabularx}
\usepackage{epstopdf}
\usepackage{mathrsfs}

\usepackage[colorlinks,linkcolor=blue,urlcolor=blue,anchorcolor=blue,citecolor=blue]{hyperref}
\newcommand{\seq}{\begin{subequations}}
\newcommand{\sen}{\end{subequations}}
\newcommand{\eq}{\begin{eqnarray}}
\newcommand{\en}{\end{eqnarray}}

\def\shiftdown#1{#1\llap{\lower.04ex\hbox{#1}}}

\journal{Nucl. Phys. A. Published in Nucl. Phys. A 968 (2017) 14--22, by C.~Huang and B.-Q.~Ma.~~}

\begin{document}

\title{Transverse charge densities of the deuteron in soft-wall AdS/QCD}

\author[PKU]{Chao Huang}

\author[PKU,CIC,CHEP]{Bo-Qiang Ma\corref{cor1}}
\ead{mabq@pku.edu.cn}
\cortext[cor1]{Corresponding author at:School of Physics,Peking University,Beijing 100871,China.}

\address[PKU]{School of Physics and State Key Laboratory of Nuclear Physics and
Technology, Peking University, Beijing 100871,
China}
\address[CIC]{Collaborative Innovation Center of Quantum Matter, Beijing, China}
\address[CHEP]{Center for High Energy Physics, Peking University, Beijing 100871, China}

\begin{abstract}

We investigate the transverse charge densities of the deuteron in a soft-wall AdS/QCD model by considering both the unpolarized and the transversely polarized cases. The deuteron form factors are derived from the soft-wall AdS/QCD model and it is shown that the calculated results agree with the  phenomenological parametrization and experimental data. The transverse charge densities of the deuteron are also derived from the soft-wall AdS/QCD model, and we notice slight deviations from those of the phenomenological parametrization.

\end{abstract}


\date{\today}


\maketitle

\section{Introduction}
The electromagnetic forms factors contain information about the internal structure of the nucleon and the nucleus.
With the Fourier transformation, form factors of nucleons reflect the spatial distributions such as charge densities~\cite{Carlson:2008hu}.
There have been considerable efforts to investigate the transverse charge densities of the nucleon~\cite{Mondal:2016,Mondal:2016nk,Chakrabarti:2014sv,Tiator:2009}.
The initial and final states for calculating form factors are different and the three dimensional Fourier transformation can not be interpreted as densities. But the transverse densities, which defined in fixed light-front time, have density interpretation~\cite{Miller:2009hh,Venkat:2011}.
Transverse charge densities are a new tool for analyzing electromagnetic form factors of systems composed of constituents that move relativistically~\cite{Miller:2010ud}.
It has led to some very interesting findings by use of transverse charge densities. For example, the center of the neutron charge densities has a negative core~\cite{Miller:2007} and the spatial extent of the magnetization density of the proton is greater than that of its charge density~\cite{Miller:2008gr}.

In recent years, anti-de Sitter (AdS) spacetime and quantum chromodynamics (QCD) correspondence, which has emerged as one of the most promising techniques to investigate the structure of hadron, has achieved significant progress in research of non-perturbative QCD.
According to the AdS and conformal field theories (CFT) conjecture~\cite{Maldacena:1998}, a weakly coupled gravity theory in $\mathrm{AdS}_{n+1}$ can be related to a conformal theory in $n$-dimensional space-time.
To apply AdS/CFT to QCD, which is not a conformal theory, the conformal invariance needs to be broken. Usually, there are two methods, referred to as the hard-wall model~\cite{Polchinski:2002ij} and the soft-wall model~\cite{Karch:2006ie}, adopted to achieve this goal. In the former method, a sharp cutoff is imposed at large distance, so that the wave functions vanish at the boundary. In the latter method, the conformal invariance is broken by a dilaton background, which provides a smooth cutoff at large distance. Compared to the hard-wall model, soft-wall AdS/QCD can reproduce the Regge trajectory~\cite{Brodsky:2009in} and the massless pion in the chiral limit~\cite{Brodsky:2014ev}.
So far, AdS/QCD has been successfully applied to describe many hadron properties such as hadron mass spectrum, generalized parton distributions, meson and nucleon form factors, transverse densities, and structure functions etc.~\cite{Vega:2011ej,Brodsky:2007,Brodsky:2006hi,Ahn:2009bc,Chakrabarti:2013ld,Brodsky:1992je,
Liu:2015qn,Guo:2016pm,Gutsche:2012ox,Brodsky:2008ud,Chakrabarti:2016dv}.

The transverse charge densities of the nucleon in a model independent way have been studied in Ref.~\cite{Miller:2007} and  the charge densities in the transverse plane for a transversely polarized nucleon are shown in Refs.~\cite{Selyugin:2009je,Carlson:2008hu}. In Ref.~\cite{Granados:2014ej}, by using the method of dispersion analysis and chiral effective field theory, transverse charge densities have been studied in the chiral periphery of the nucleon. The transverse densities for the quarks of the nucleon are analyzed in a chiral quark-soliton model~\cite{Miller:2010ud}. Apart from the transverse charge densities of the nucleon, it is also  interesting to study the transverse densities of the deuteron. For example, the transverse charge densities of the deuteron have been studied in Ref.~\cite{Carlson:2009mv}, by using the parametrization of the deuteron form factor data given as fit~\uppercase\expandafter{\romannumeral2} by Abbott et al.~\cite{Abbott:2000ak}.

In this work, we give a comparison of the deuteron form factors calculated in the soft-wall model with those from the experimental data and phenomenological parametrization. Then, we give a prediction for the transverse charge densities of the deuteron in the soft-wall AdS/QCD model and compare the results with the parametrization.


The paper is organized as follows. The electromagnetic form factors of the deuteron in the soft-wall AdS/QCD model are given in Sec.~\uppercase\expandafter{\romannumeral2}. In Sec.~\uppercase\expandafter{\romannumeral3}, the transverse charge densities of the deuteron for both unpolarized and transversely polarized cases are discussed. Finally, we provide a brief summary in Sec.~\uppercase\expandafter{\romannumeral4}.

\section{SOFT-WALL ADS/QCD MODEL FOR DEUTERON FORM FACTORS}
\begin{figure}
 	\begin{center}
 		\includegraphics[width=0.45\textwidth]{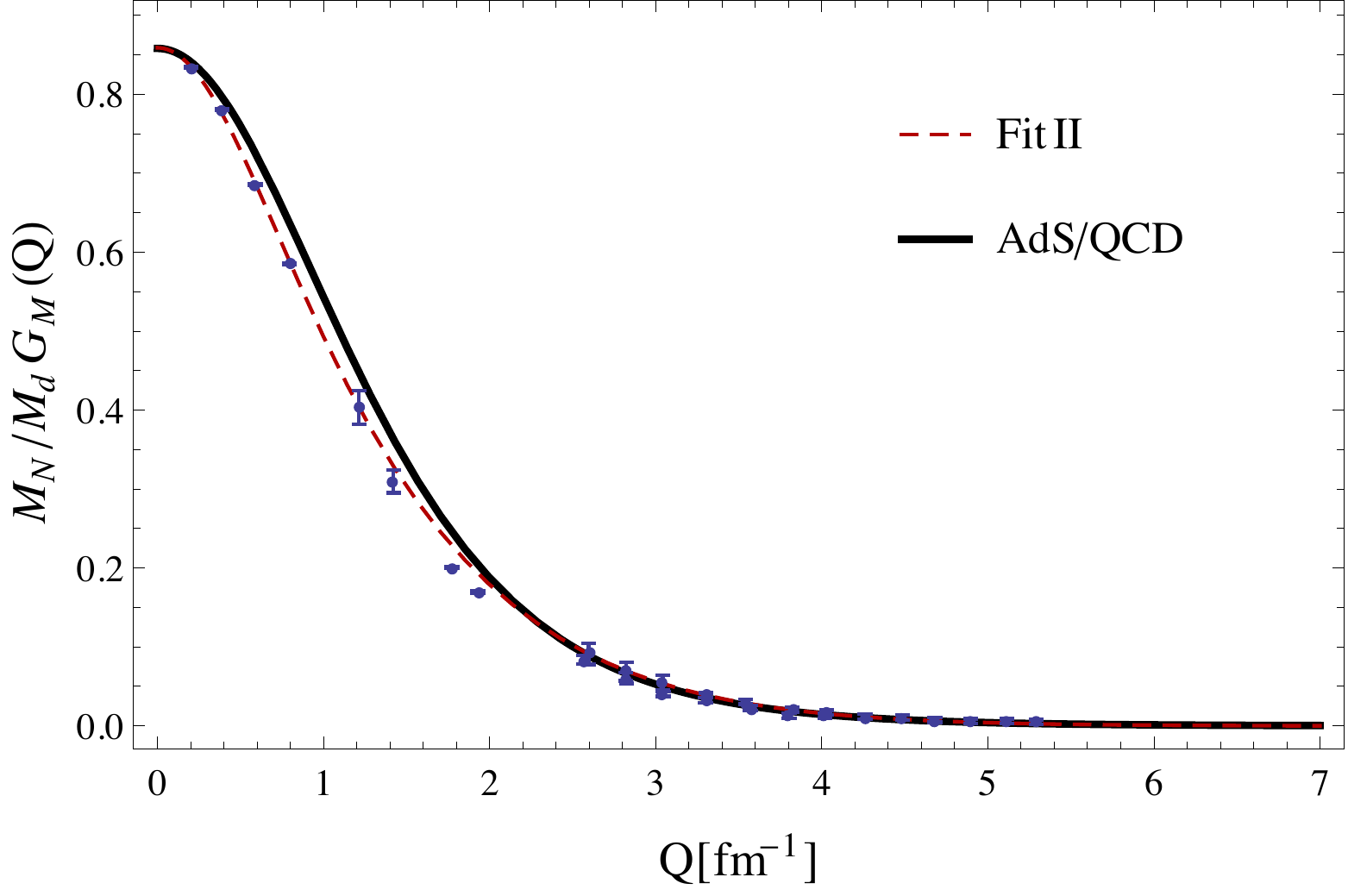}
 	\end{center}
 	\vspace{-0.5cm}
 	\caption{The plot shows the magnetic deuteron form factor $G_M$. The solid black curves are for the soft-wall AdS/QCD model~\cite{Gutsche:2015qh}, and the red dashed curves represent the JLab t20 Fit \uppercase\expandafter{\romannumeral2}~\cite{Abbott:2000ak}. The experimental data are from~\cite{Abbott:2000ak,Holt:2012gg}.}
 \end{figure}

\begin{figure*}[htbp]
   \includegraphics[width=0.45\textwidth]{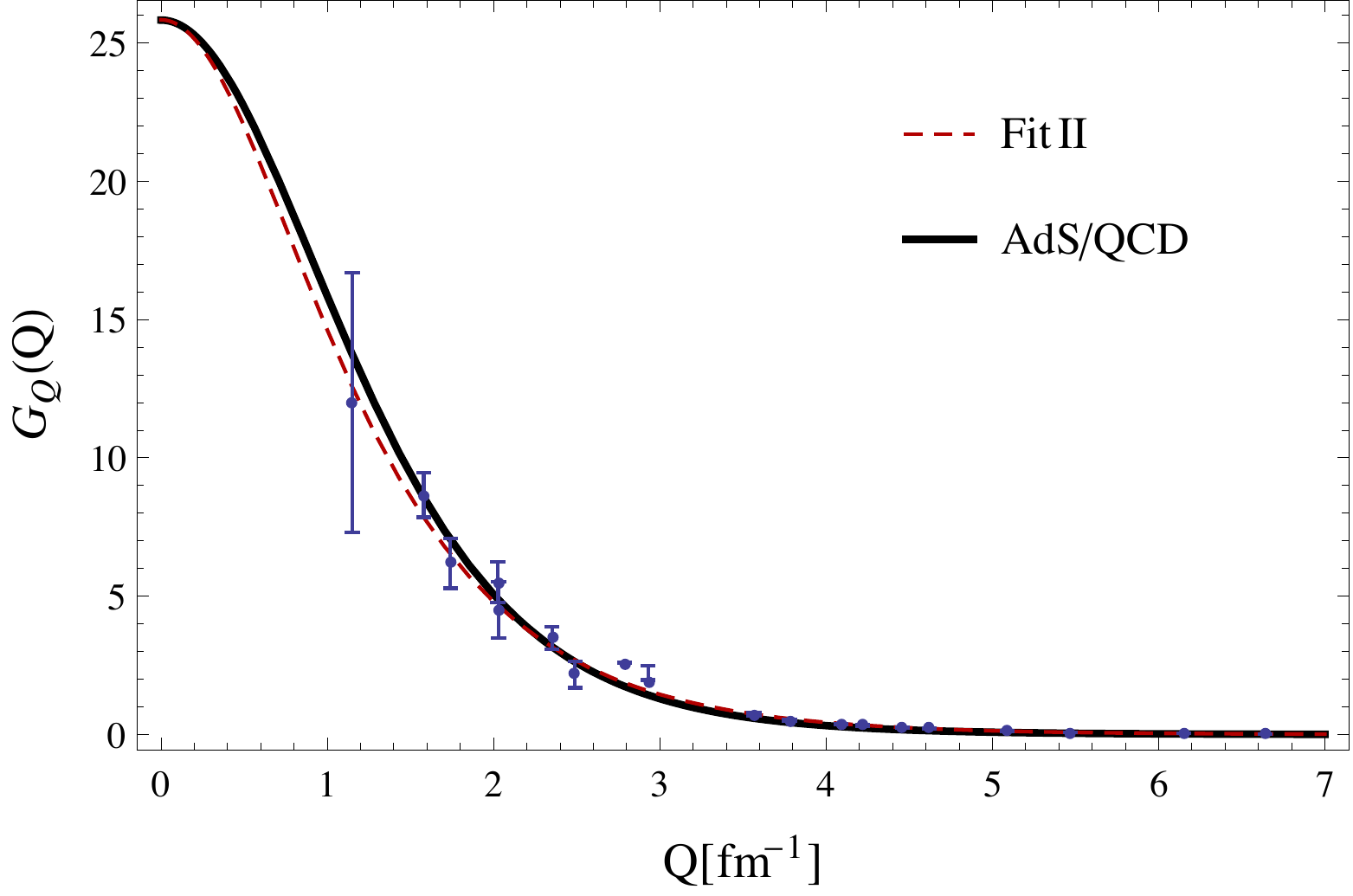}
   \includegraphics[width=0.45\textwidth]{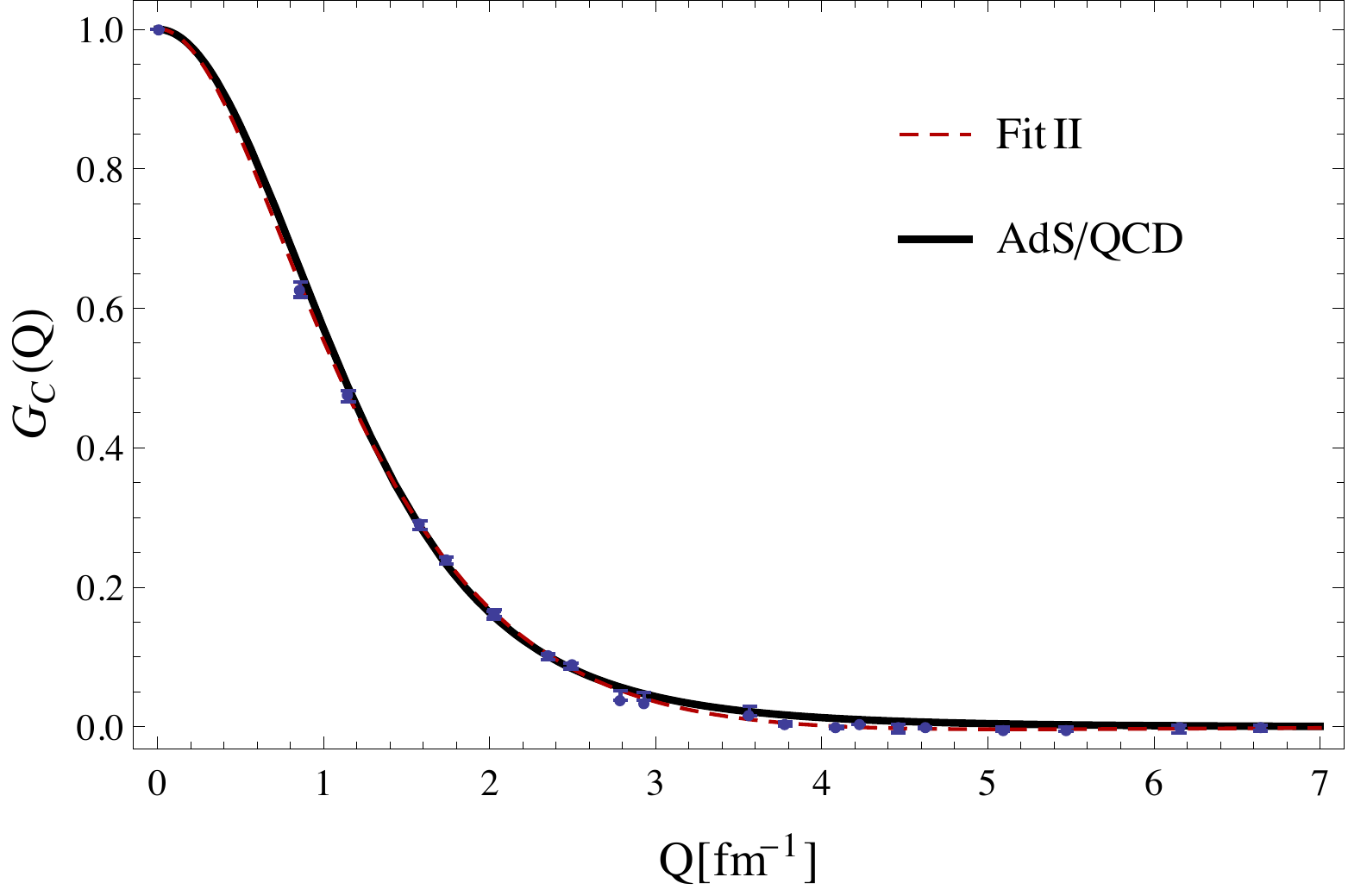}
\caption{\label{Sachs_ratio} The plots show the charge and quadrupole  form factors $G_C$ and $G_Q$ for the deuteron between the soft-wall AdS/QCD model and the parametrization. All the notations are same as Fig.~1.}
\end{figure*}

For the deuteron electromagnetic form factors, we consider the soft-wall model of AdS/QCD proposed by Gutsche et al.~\cite{Gutsche:2015qh,Gutsche:2016id}. The effective action is given by
\begin{equation}
\begin{split}
S =
  & \int d^4xdz \sqrt{g} e^{-\varphi(z)} \biggl[ - \frac{1}{4} F_{MN} F^{MN} - D^M d^\dagger_{N}  D_M d^N - i c_2(z) F^{MN} d^\dagger_{M} d_{N}\\
  &  + \frac{c_3(z)}{4M_d^2} \, e^{2A(z)} \, \partial^M F^{NK} \biggl( iD_K d^\dagger_{M} d_{N} - d^\dagger_{M} i D_K d_{N} + \mathrm{H.c.} \biggr) + d^\dagger_{M} \, \Big(\mu^2 +  U(z) \Big) \times d^M \biggr] \, ,\\
\end{split}
\end{equation}
where $g=|\mathrm{det} g_{MN}| = e^{10A(z)}$. $d^M(x,z)$ and $V^M(x,z)$ are AdS fields which are dual to the Fock compoment contributing to the deuteron with twist $\tau = 6$ and the electromagnetic field respectively. $A(z)$ is equal to $\log(R/z)$ and $F^{MN}(x,z)=\partial^M V^N (x,z)-\partial^N V^M(x,z)$ is the stress tensor. The covariant derivative is $D^M = \partial^M - i e V^M(x,z)$, $\mu^2 R^2 = (\Delta -1)(\Delta -3)$ is five-dimensional mass, $R$ is the AdS radius, $\varphi(z) = \kappa^2 z^2$ is the background dilaton field, $\Delta = \tau + 1$ is the dimensional of the $d^M(x,z)$ field, and $M_d$ is the deuteron mass. $U(z)=(\varphi(z)/R^2) U_0$ is the confinement potential, where $U_0$ is fixed by the value of the deuteron mass.
The $z$-dependent couplings are $c_2(z) = e^{-\beta \varphi(z)} [c_2^{(1)} + c_2^{(2)} e^{\alpha_2 \log \varphi(z)}]$ and $c_3(z) =c_3 e^{-\beta \varphi(z) + \alpha_3 \log \varphi(z)}$, where the couplings are fixed from the normalization of $G_i$ and the fitting to data.


In this method, the deuteron is described as a vector field with a gauge field and only the mass of the deuteron is contained in the effective action. We simply treat the deuteron as an effective proton-neutron bound state of six quarks, or an active quark and an effective spectator in analogy to the AdS/QCD model of baryons~\cite{Liu:2015qn}.
The parameter $\kappa$ of the deuteron is smaller than that of the nucleon and this difference can be related to the different sizes of the deuteron and the nucleon. On the other hand, the parameters of different particles with different numbers of quarks can be different in different situations of mesons and beryons~\cite{Wang:2016wo}.

The confinement potential is usually derived from the deformation of the AdS space by introducing a dilaton $\varphi = \kappa^2 z^2$~\cite{Liu:2015qn}. The specific form of the confinement potential is used to produce the correct 
asymptotic behaviors of the form factors of the deuteron at high $Q^2$.

The gauge-invariant matrix element describing the interaction of the deuteron with the external vector field is~\cite{Gutsche:2015qh}
\begin{equation}
\begin{split}
M_{\mathrm{inv}}^{\mu} (p,p') =
   &-\Biggl[ G_1(Q^2) \epsilon^* (p')\cdot\epsilon(p) - \frac{G_3(Q^2)}{2M^2_d} \epsilon^* (p') \cdot q \epsilon(p) \cdot q \biggr] (p+p')^{\mu} \\
   &   -G_2(Q^2)\left[ \epsilon^{\mu} (p)\cdot\epsilon(p')  \right. \left. - \epsilon^{* \mu} (p') \cdot q \epsilon(p) \cdot q \right] ,\\
\end{split}
\end{equation}
where $\epsilon~(\epsilon^*)$ and $p~(p')$ are the polarization and four-moment of the initial~(final) deuteron. $q = p' - p$ is the 4-momentum transfer and $Q^2 = - q^2 $ .

\begin{figure*}[htbp]
\begin{minipage}[c]{0.98\textwidth}
\small{(a)}
\includegraphics[width=0.45\textwidth]{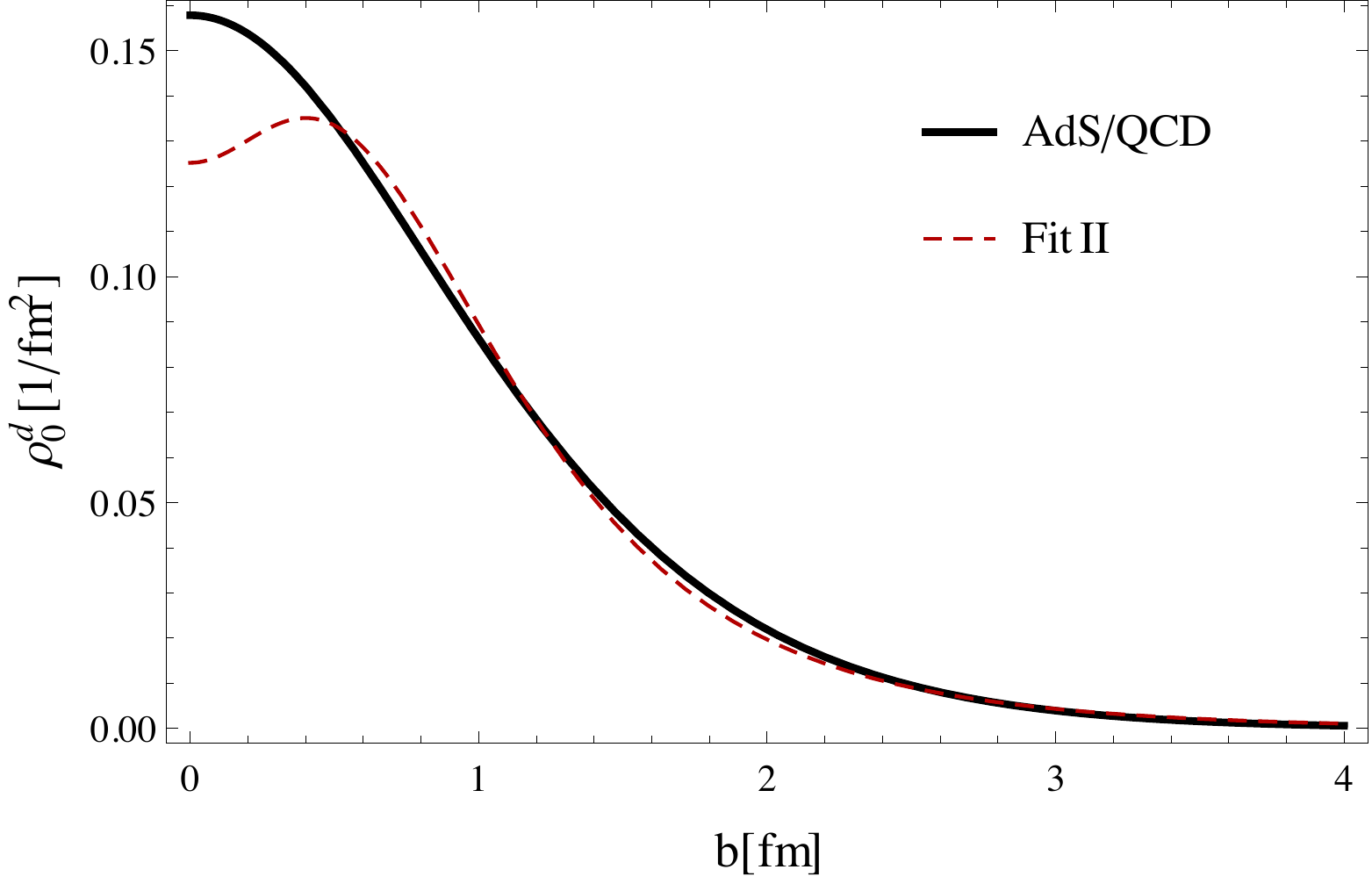}
\hspace{0.1cm}%
\small{(b)}\includegraphics[width=0.45\textwidth]{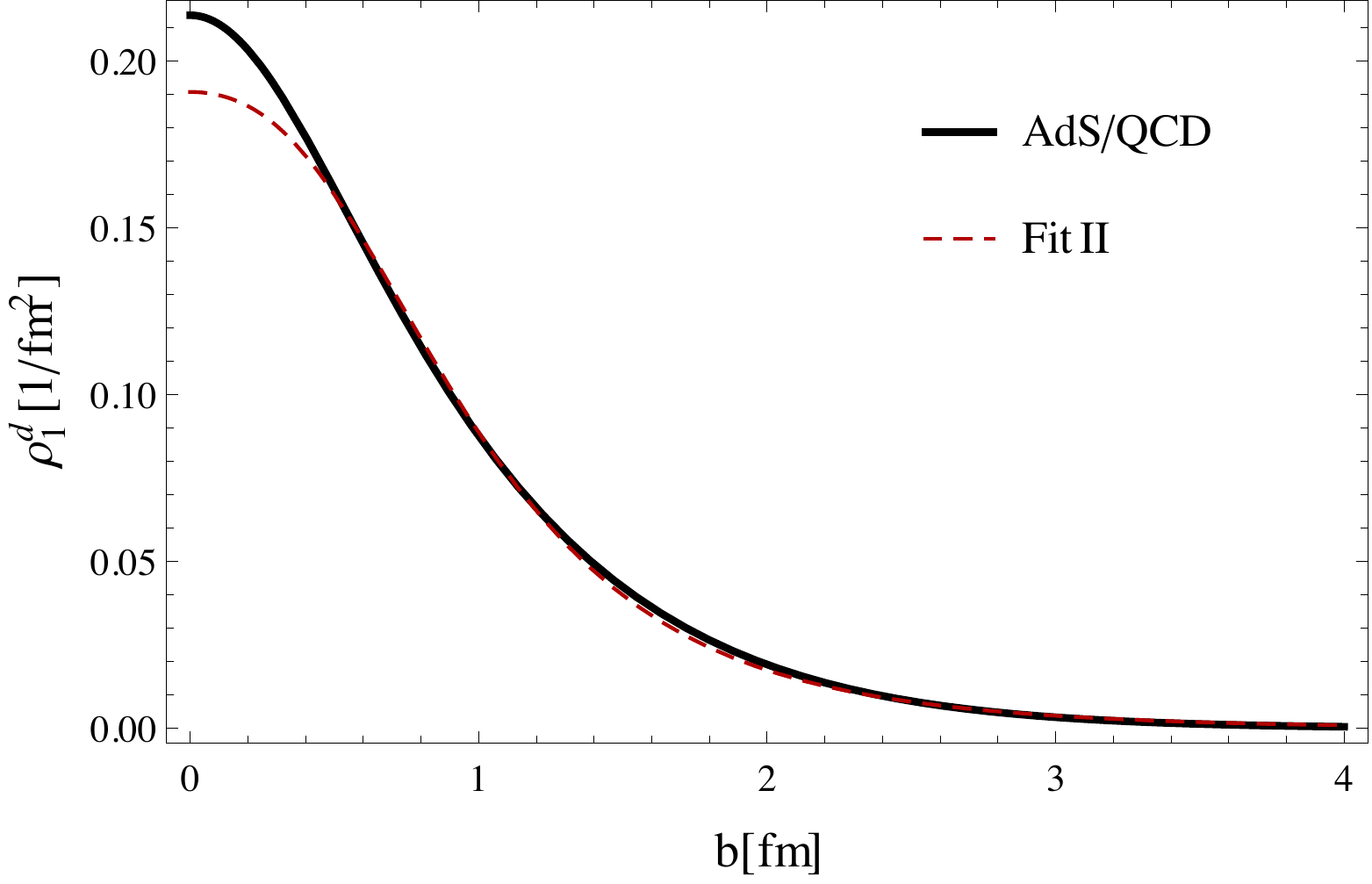}
\end{minipage}
\begin{minipage}[c]{0.98\textwidth}
\small{(c)}
\includegraphics[width=0.45\textwidth]{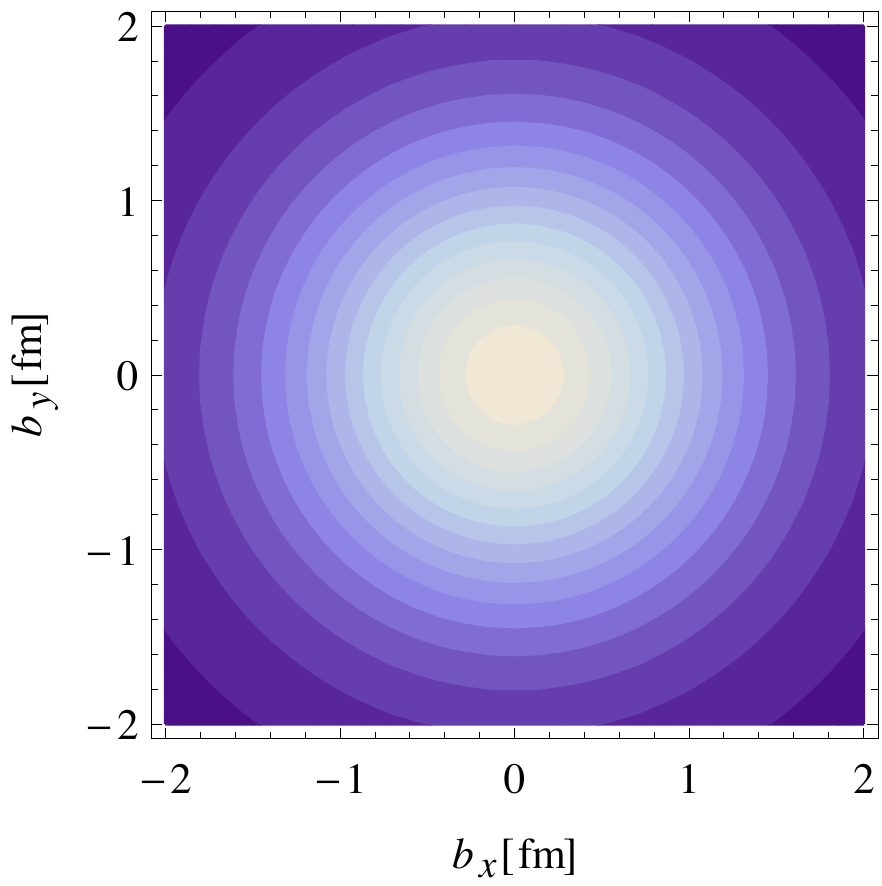}
\hspace{0.1cm}%
\small{(d)}\includegraphics[width=0.45\textwidth]{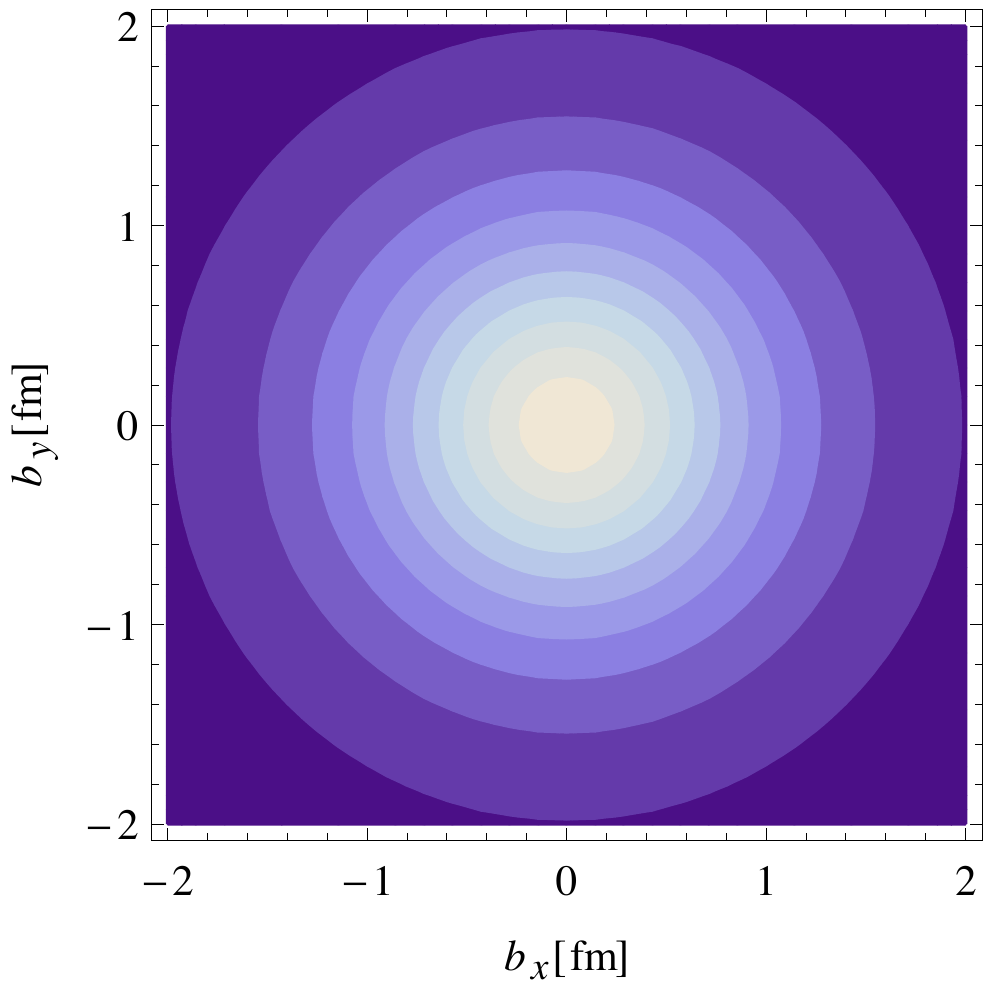}
\end{minipage}
\caption{Transverse charge densities for the unpolarized deuteron.(a) and (c) represent $\rho^d_0$, and (c) is a top view of three-dimensional charge density. (b) and (d) represent $\rho^d_1$.}
\end{figure*}

The three electromagnetic form factors $G_{1,2,3}$ of the deuteron are related to the charge $G_C$, quadrupole $G_Q$ and magnetic $G_M$ form factors by~\cite{Gilman:2002dv}
\begin{equation}
\begin{split}
&G_C = G_1 + \frac{2}{3} \tau_d G_Q , \\
&G_M = G_2, \\
&G_Q = G_1 - G_2 + (1+\tau_d) G_3, \\
\end{split}
\end{equation}
where $ \tau_d = {Q^2}/{4M^2_d}$ and the form factors $G_{C,M,Q}$ are all functions of $Q^2$.

The form factors are normalized such that~\cite{Garcon:2001ud}
\begin{equation}
\begin{split}
&G_C(0) = 1 , \\
&G_M (0)= M_d^2 \mathscr{Q_d} , \\
&G_Q (0)= \frac{M_d}{M_N} \mu_d, \\
\end{split}
\end{equation}
where $\mathscr{Q_d}$ and $\mu_d$ are the quadrupole and magnetic moments of the deuteron.

Thus, the form factors for the deuteron in this model are given by~\cite{Gutsche:2016id}
\begin{equation}
\begin{split}
G_1 (Q^2) = & \frac{\Gamma(a+1) \Gamma(7)}{\Gamma(a+7)}, \\
G_2 (Q^2) = & c_2^{(1)} \frac{I_1(Q^2)}{I_1(0)} + c_2^{(2)} \frac{I_2(Q^2)}{I_2^{(0)}} , \\
G_3 (Q^2) = & c_3 \frac{I_3(Q^2)}{I_3(0)}, \\
\end{split}
\end{equation}
where
\begin{equation}
\begin{split}
I_i (Q^2) = & \frac{ \Gamma (7+\Delta_i) }{\Gamma(6)}  \int_{0}^{1} d x x^a \frac{(1-x)^{5+\Delta_i}}{(1+\beta (1-x))^{7+ \Delta_i } }, \\
\Delta_1 =& 0 , \quad \Delta_2 = \alpha_2, \\
\Delta_3 =& \alpha_3, \quad  c_2^{(1)}=2I_1(0). \\
\end{split}
\end{equation}

In this model, we have two free parameters: $\kappa$ and $U_0$. The latter one is fixed by the deuteron mass, and the formal one is used to fit the form factors of the deuteron. In this work, $\kappa = 0.19~\mathrm{GeV}$ is adopted to have a best fit to three form factors.
The asymptotic behaviors of the deuteron form factors are $G_M (Q^2) \sim G_Q(Q^2) \sim 1/Q^{12}$ and $G_C(Q^2) \sim 1/Q^{10} $. The asymptotic behaviors of $G_i$ are at high $Q$, and the values of the form factors are very small at high $Q$, therefore, the difference between $1/Q^{12}$ and $1/Q^{10}$ is small at high $Q$ with small form factors. Such behaviors are consistent with the prediction of perturbative QCD and the relation proposed in Ref.~\cite{Brodsky:1992je}.
The parameter $\kappa$ of the deuteron is smaller than that of the nucleon. This difference can be related to the fact that the size of the deuteron is bigger than that of the nucleon. In fact, the parameters of different particles with different numbers of quarks can be different~\cite{Liu:2015qn, Guo:2016pm, Wang:2016wo}.

In Fig.~1 and Fig.~2, we show the charge, quadrupole and magnetic form factors ($G_{C,Q,M}$) for the deuteron calculated in the framework of the soft-wall AdS/QCD model and compare the results with those from the global parametrization~\cite{Abbott:2000ak}.
The plots suggest that for the charge deuteron form factor $G_C$ the model and the phenomenological parametrization both agree with the experimental result well. For the quadrupole and magnetic deuteron form factors, the overall description of the soft-wall AdS/QCD model is consistent with the experimental data and the parametrization. But around $Q=1~\mathrm{fm}^{-1}$, there is a slight deviation.

\section{TRANSVERSE CHARGE DENSITIES}

The expression of transverse charge densities for an unpolarized deuteron is given by the helicity form factors $G^+_{\lambda \lambda}$~\cite{Carlson:2009mv},
\begin{eqnarray}
\rho^d_{\lambda} (b)
\!\!&\equiv& \!\!\int \frac{d^2 \vec q_\perp}{(2 \pi)^2}
e^{- i \, \vec q_\perp \cdot \vec b}  \frac{1}{2 P^+}
\langle P^+, \frac{\vec q_\perp}{2}, \lambda
| J^+ | P^+, \frac{- \vec q_\perp}{2}, \lambda  \rangle
\nonumber \\
\!\!&=&\!\! \int_0^\infty \frac{d Q}{2 \pi} Q \,
J_0(b \, Q) \, G^+_{\lambda \lambda}(Q^2),
\end{eqnarray}
where ${\bf b} = b(\cos \phi_b \hat{e}_x + \sin \phi_b \hat{e}_y)$ represents the position from the transverse center of mass of the deuteron, and $J_n$ is the cylindrical Bessel function of order $n$.
\begin{figure*}[htbp]
\begin{minipage}[c]{0.98\textwidth}
\small{(a)}
\includegraphics[width=0.45\textwidth]{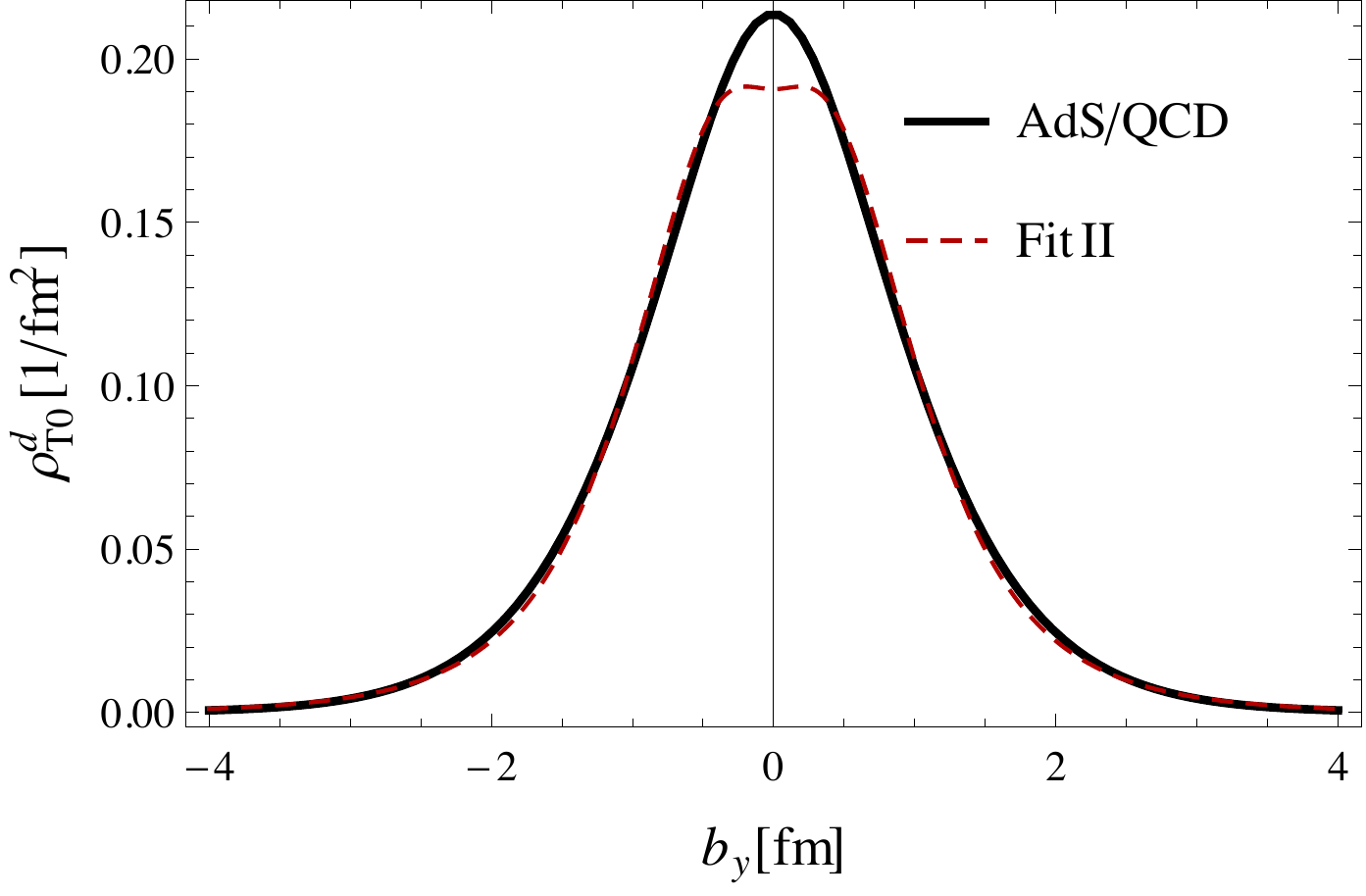}
\hspace{0.1cm}%
\small{(b)}\includegraphics[width=0.45\textwidth]{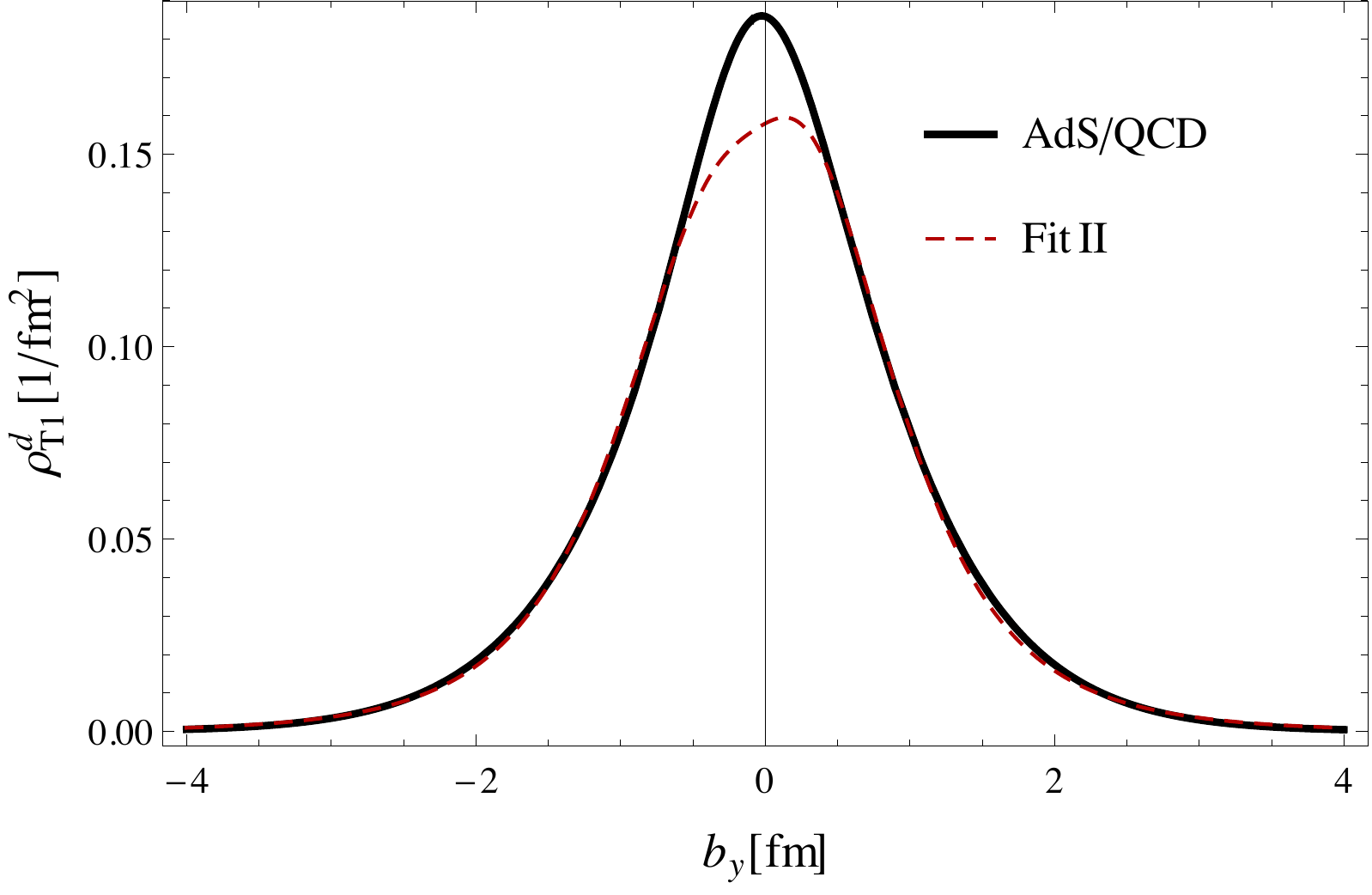}
\end{minipage}
\begin{minipage}[c]{0.98\textwidth}
\small{(c)}
\includegraphics[width=0.45\textwidth]{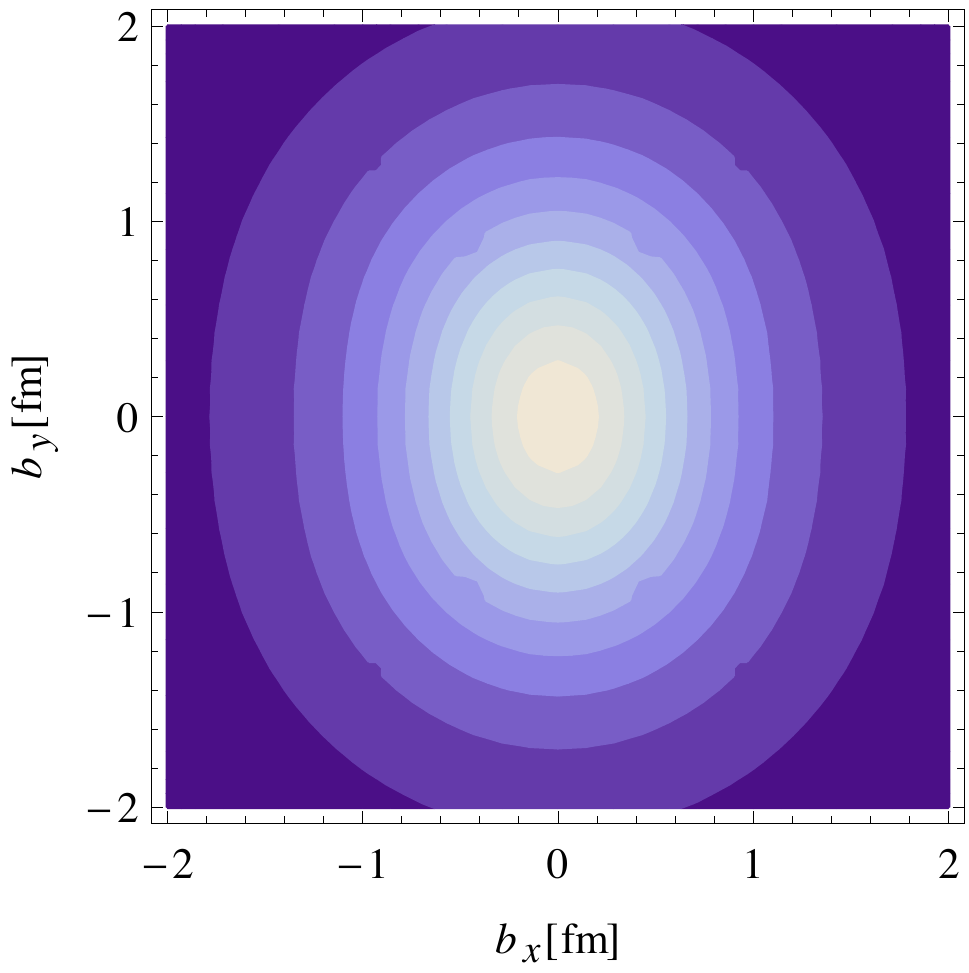}
\hspace{0.1cm}%
\small{(d)}\includegraphics[width=0.45\textwidth]{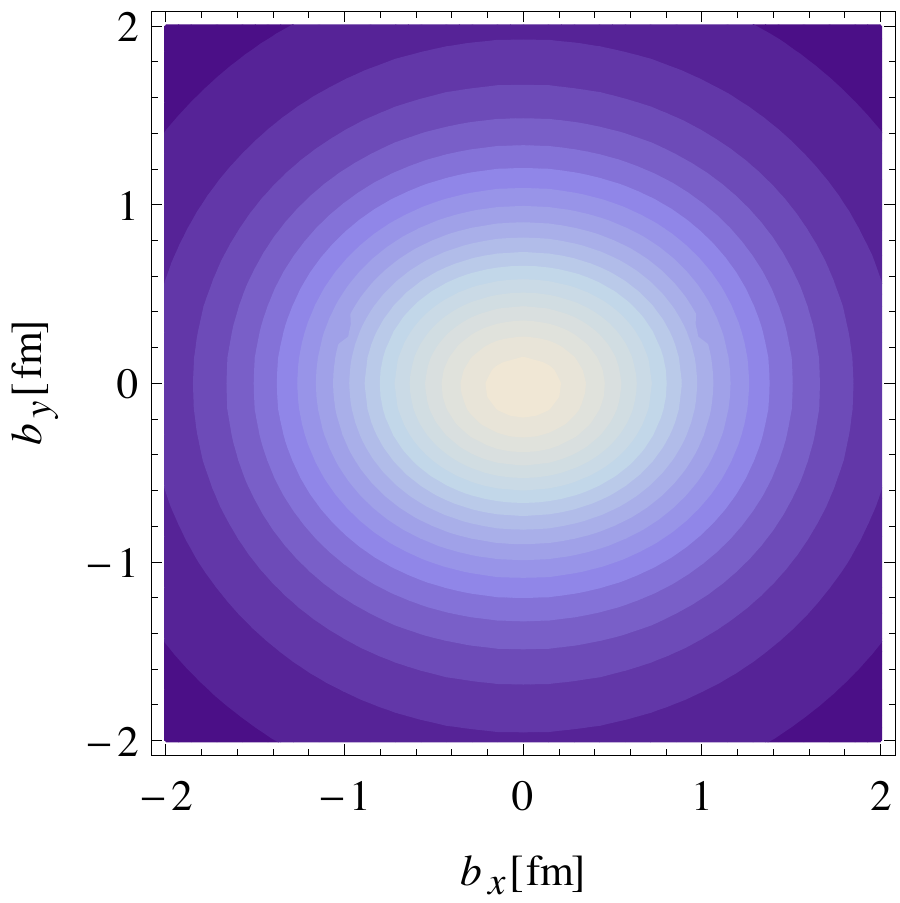}
\end{minipage}
\caption{Transverse charge densities for the polarized deuteron.(a) and (c) represent $\rho^d_{T0}$, and (c) is a top view of three-dimensional charge density. (b) and (d) represent $\rho^d_{T1}$. The polarization of the deuteron is along the $x$ direction.}
\end{figure*}
$\lambda = \pm 1,0$ $(\lambda' = \pm 1,0)$ denotes the initial (final) deuteron light front helicity, and the helicity form factors $G^+_{\lambda \lambda'}$ are real and depend only on $Q^2$. Note that $\rho^d_{\lambda}$ only depends on $b=| \vec{b} |$. It is well defined for all values of $b$, even $b$ is smaller than the Compton wavelength~\cite{Carlson:2008hu}. The helicity-conserving form factors $G^+_{11}$ and $G^+_{00}$ can be expressed in terms of $G_{C,M,Q}$ as
\begin{equation}
\begin{split}
G^+_{11} =& \frac{1}{1+\tau_d} \left[ G_C +\tau_d G_M +\frac{\tau_d}{3} G_Q \right], \\
G^+_{00} =& \frac{1}{1+\tau_d} \left[ (1- \tau_d) G_C +2\tau_d G_M - \frac{2\tau_d}{3} (1+2\tau_d) G_Q \right].\\
\end{split}
\end{equation}

We show the transverse charge densities for the unpolarized deuteron in Fig.~3(a) and Fig.~3(b). The plots imply that the predictions of the soft-wall AdS/QCD model for the unpolarized transverse densities are in good agreement with the global parametrization at $b > 0.6~\mathrm{fm}$. But around the center of mass ($b=0$), the soft-wall AdS/QCD prediction shows a little higher value of the transverse charge densities compared to the parametrization, especially $\rho^d_0$. The top view of three dimensional transverse charge densities for unpolarized deuteron are shown in Fig.~3(c) and Fig.~3(d). From the plots, we can notice that the unpolarized transverse charge densities are axially symmetric. Compared to the phenomenological parametrization, which finds a dip in Fig.~3(a), the model calculation does not have a dip in the center of the three dimensional transverse charge density. As emphasized in~\cite{Carlson:2009mv}, the dip seen in the figure of the phenomenological parametrization is based on observation and a light front interpretation of the data. So, in order to give a more convinced result, the more accurate experimental data of the deuteron form factors are needed.


For a transversely polarized deuteron, we suppose that the transverse polarization direction is $S_{\perp} = \cos \phi_S \hat{e}_x + \sin \phi_S \hat{e}_y$. The transverse charge densities can be defined as
\begin{equation}
\begin{split}
\rho^d_{T s_{\perp}} ({\bf b}) =
    &\int \frac{d^2 {\bf q}_{\perp} }{(2\pi)^2} e^{-i {\bf q}_{\perp} \cdot {\bf b}} \frac{1}{2P^+}  \\
    &\times \langle P^+, \frac{{\bf q}_{\perp}}{2} , s_{\perp} | J^+ | P^+ , \frac{- {\bf q}_{\perp}}{2} ,s_{\perp} \rangle ,\\
\end{split}
\end{equation}
where $s_{\perp}$ is the deuteron spin projection along the direction of ${\bf S}_{\perp}$.

By using the Fourier transformation, one gets~\cite{Carlson:2009mv}
\begin{eqnarray}
\rho^d_{T \, 1} (\vec b) & = & \int_0^\infty \frac{d Q}{2 \pi} \,
Q\, \left\{
J_0(b \, Q) \, \frac{1}{2} \left( G^+_{1 \, 1} + G^+_{0 \, 0} \right)
\right. \nonumber \\
&&\hspace{0.5cm}
+ \sin(\phi_b - \phi_S) \, J_1(b \, Q) \sqrt{2} \,  G^+_{0 \, 1}  \\ \nonumber
&&\left. \hspace{0.5cm}- \cos 2 (\phi_b - \phi_S) \, J_2(b \, Q)
\frac{1}{2} \, G^+_{-1 \, +1}
\right\} , \\
\rho^d_{T \, 0} (\vec b) & = & \int_0^\infty \frac{d Q}{2 \pi} \,
Q\, \left\{ J_0(b \, Q) \, G^+_{1 \, 1} \right. \nonumber \\
&&\left. \hspace{0.5cm}+ \cos 2 (\phi_b - \phi_S) \, J_2(b \, Q)
\, G^+_{-1 \, +1} \right\} ,
\end{eqnarray}
where the deuteron helicity form factors can be expressed in terms of $G_{C,Q,M}$ as~\cite{Carlson:2009mv}
\begin{equation}
\begin{split}
&G^+_{01} = - \frac{\sqrt{2 \eta }}{1+\eta} \left[ G_C - \frac{1}{2}(1-\eta) G_M +\frac{\eta}{3} G_Q \right], \\
&G^+_{-1 +1} = \frac{\eta}{1+\eta} \left[  G_C - G_M -(1+\frac{2\eta}{3})G_Q \right].\\
\end{split}
\end{equation}

Without loss of generality, the polarization of the deuteron along the $x$ axis (i.e., for $\phi_s = 0$) is taken in this work. In Fig.~4(a) and Fig.~4(b), we show the transverse charge densities for polarized deuteron using the form factors obtained in both soft-wall AdS/QCD model and phenomenological parametrization. In Fig.~4(c) and Fig.~4(d), we draw a top view of three-dimensional charge densities for deuteron polarized along the $x$ direction.
The soft-wall AdS/QCD model is in good agreement with the parametrization, except that it provides larger peak densities than the parametrization.
From the plots, we can see that the transverse charge density $\rho^d_{T1}$ for the transversely polarized deuteron gets displaced toward the $b_y$ direction, while so does $\rho^d_{T1}$ toward the $b_x$ direction.

\section{SUMMARY}

In this paper, we present a detailed study of the transverse charge densities for deuteron in a soft-wall AdS/QCD model. The results are compared with the phenomenological parametrization of the form factors. Both the unpolarized and transversely polarized cases are considered in this work. The results show that the deuteron form factors calculated in the soft-wall AdS/QCD model and the parametrization are in good agreement with each other. The transverse charge densities are studied in detail in the soft-wall AdS/QCD model, and the results are compared with the parametrization. The results calculated in the two methods are consistent with each other, except that there are some slight deviations at the center of mass ($b = 0$). We suggest that in order to test these different predictions, more accurate experiments of deuteron form factors are needed in future.

\section*{Acknowledgments}
This work is supported by the National Natural Science Foundation of China (Grants No.~11475006 and  No.~11120101004).

\end{document}